\documentclass[12pt]{article}
\usepackage{amsmath,amsfonts,amssymb,graphicx,hyperref}

\begin{document}

\title{Control of the Extension-Flexion Cycle of Human Knees During Bicycle Riding by a Synergy of Solitary Muscular Excitations and Contractions}
\author{Zoran Gojkovic\thanks{Clinical Centre Vojvodina, Serbia; e-mail: gojkoz@gmail.com} ~~and ~Tijana Ivancevic\thanks{Tesla Science Evolution Institute, Adelaide, Australia; e-mail: tijana.ivancevic@alumni.adelaide.edu.au}}
\date{}
\maketitle

\vspace{2cm}
\begin{abstract}
A Hill-type model is proposed for the extension-flexion cycle of human knees during bicycle riding. The extension-flexion cycle is controlled by a synergy of muscular excitations and contractions of the knee musculature. Muscular action potentials are modeled by Sine-Gordon kinks, while titin-influenced actomyosin contractions are modeled by Korteweg-de Vries solitons. As an application, the total knee arthroplasty is discussed. \bigbreak\bigbreak

\end{abstract}

\newpage
\tableofcontents

\vspace{2cm}
\section{Introduction}

Human knee joint has a complex anatomical structure (including three articulations: tibio-femoral, tibio-fibular and patelo-femoral) and a highly nonlinear biomechanical function (see Figure 3.20 in \cite{TijSpr14} and the related references, as well as \cite{Gojko1}). Its primary dynamical function is a single dominant degree-of-freedom (DOF): flexion/extension in the sagittal plane, actuated by the two of the largest muscular groups in the human body: m. quadriceps femoris performing knee extension and m. hamstrings performing knee flexion.\footnote{In reality, human knee is much more flexible than the humanoid robot knee: it also allows a restricted latero-medial rotation in the semi-flexed position (which allows voluntary steering in alpine skiing), as well as three micro-translations in the tibio-femoral, tibio-fibular and patelo-femoral articulations, which are the primary cause of the knee injuries - when they become macro-translations \cite{ivGenInjury}.}

Modelling and simulation studies of the knee dynamics have usually been based on multi-body models and/or finite element method (see \cite{Machado,Weed} and the references therein). In the present paper we extend this approach by including soliton models of excitation and contraction of the extensor-quadriceps and flexor-hamstrings. We understand that, apart from muscular actuators and internal knee-joint structure, many other factors are also important for a thorough knee biomechanics, including fascia, hormonal, neural and other factors, but for the sake of simplicity we are omitting them at this stage and plan them for the larger future research.

The exercise of riding a stationary bicycle ergometer is a natural movement
performed under safe and controlled conditions, well-suited for rehabilitation in patients after the knee-replacement surgery.
Although in reality this cyclic movement includes more DOFs
than only two flexions and extensions in the knees (both hips and ankles
are also involved, together with some support from the arms and upper body),
for the purpose of this paper we focus on the two angular movements in the knees
only. Still, all the neurophysiological, spatiotemporal dynamical and
biomechanical concepts and models presented in this paper apply to the
musculature of the hips and ankles as well.

\section{Neuromuscular Background}

\subsection{Sherrington's Reciprocal Neuromuscular Excitation and Inhibition}

Classical Sherrington's\footnote{%
Nobel Laureate Sir Charles S. Sherrington was the father of the integral
neurophysiology (beyond the simple reflex arc), first published in his seminal work
\cite{Sherrington}.} theory of \emph{reciprocal neuromuscular excitation and
inhibition} can be applied to the bicycle riding movement as follows. Let us
assume that we start the bicycle by pressing-down the pedal with the left
leg. This \emph{voluntary action} is performed by \emph{excitation} and
subsequent \emph{contraction} of the knee-extensor muscles (quadriceps
femoris group) of the left leg. This voluntary ``command intent'' is
immediately followed by the following three \emph{reflex actions,} one
performed by the same (left) leg and the other two performed by the other
(right) leg. On the leading left leg, we can observe the \emph{inhibition}
and subsequent \emph{relaxation} of the knee-flexor muscles (hamstrings
group). At the same time, the right leg is automatically pulling-up the
pedal, by exciting-and-contracting its knee-flexors and simultaneously
inhibiting-and-relaxing its knee-extensors.

This is the general way how the Central Nervous System (mainly Cerebellum)
coordinates any kind of energy-efficient cyclic motion in both humans and
legged animals, like walking, running, flying, swimming, etc. If we want to
artificially enhance the patient's performance in the bicycle riding (and in
the process increase the speed of the patient's recovery) we can include the
electro-muscular stimulation (EMS) of the quadriceps and hamstrings groups
in both legs, by carefully designing the EMS-control pattern in such a way
to naturally blend into Sherrington's reciprocal neuromuscular excitation
and inhibition.

\begin{figure}[h]
\centerline{\includegraphics[width=10cm]{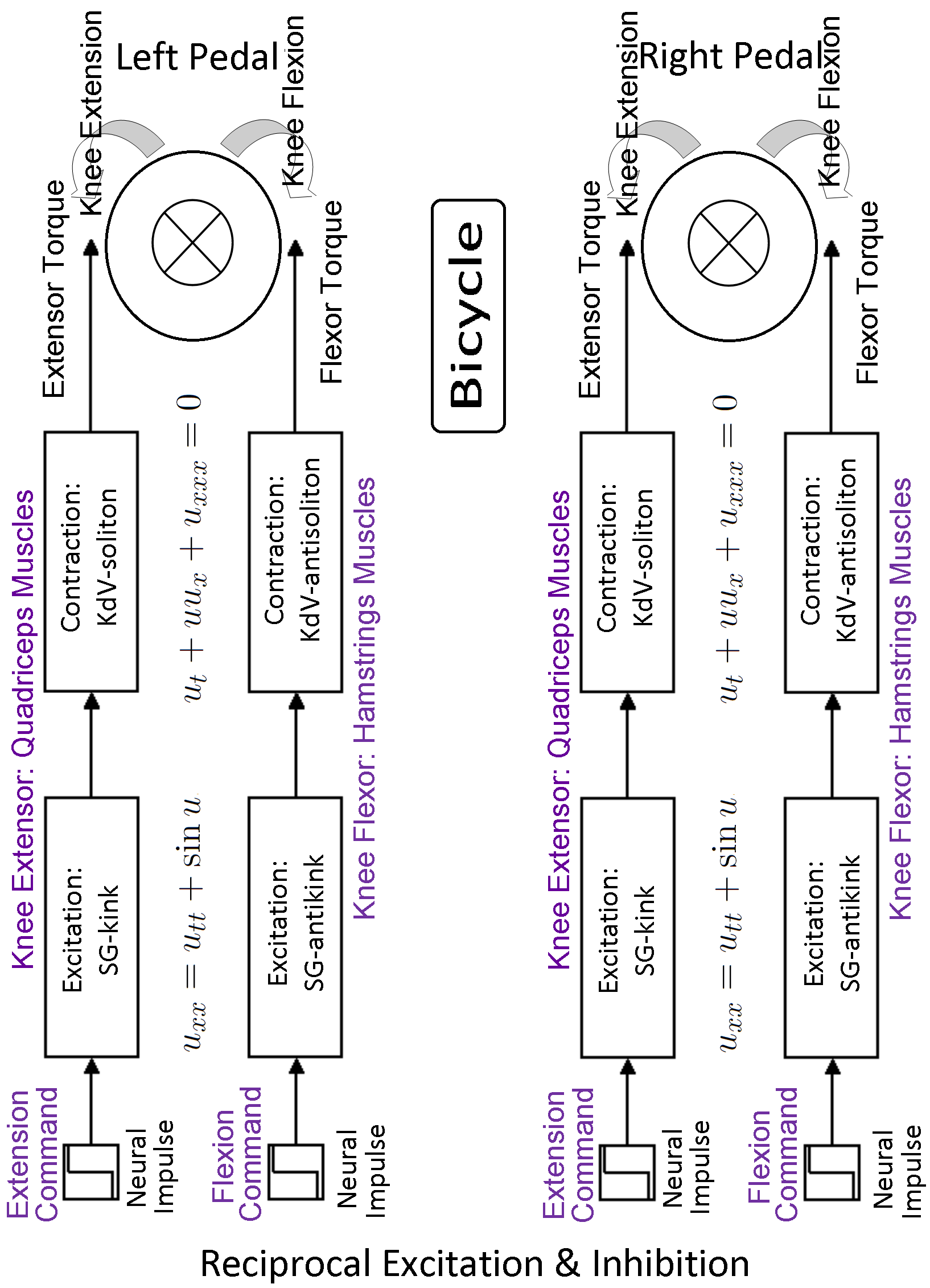}}
\caption{Schematic of the knee extension-flexion model.}
\label{GojkoKneeModel}
\end{figure}

On the nonlinear dynamics side of this problem, we recognize the need for
modelling two pairs of actions performed by the knee musculature on both
legs: excitation/inhibition and contraction/relaxation. All these
spatiotemporal actions can be modeled as appropriate solitons, which are
functions of time and muscular shortening/stretching. In this paper, we will
model the knee actions during the bicycle riding movement using solitary
excitations and contractions of the flexors and extensors of both knees (see
Figure \ref{GojkoKneeModel}).

\subsection{Sliding Filaments of Muscular Contraction: Actin, Myosin and Titin}

The field of muscular mechanics is based on two theories of muscular contraction, macroscopic and microscopic. Firstly, in 1930s, A.V. Hill developed the macroscopic contraction theory (see \cite{Hill1}) rooted in thermodynamics and resulting in his hyperbolic-shaped \emph{force-velocity relation},\footnote{Hill's force velocity relation basically states that the active muscular force is strongest in \emph{isometric} conditions, while the speed of contraction is maximal with the smallest resistance.} which is still the foundation of human movement biomechanics.
The microscopic contraction theory is called the \emph{sliding filament theory,} developed in 1954 jointly by two Huxleys \cite{HuxleyA,HuxleyH}, who proposed that the muscular contraction force is generated by the relative motion/interaction between two kinds of protein filaments, thick \emph{myosin} filaments and thin \emph{actin} filaments, biochemically fueled by the ATP-release, within the basic muscular micro-units called \emph{sarcomeres} (bounded by the \emph{Z-discs;} for a recent review, see, e.g. \cite{Krans}). This microscopic contraction theory is usually captured in the (skewed) bell-shaped \emph{force-length relation} of the \emph{concentric} muscular contraction.

\begin{figure}[h]
\centerline{\includegraphics[width=16cm]{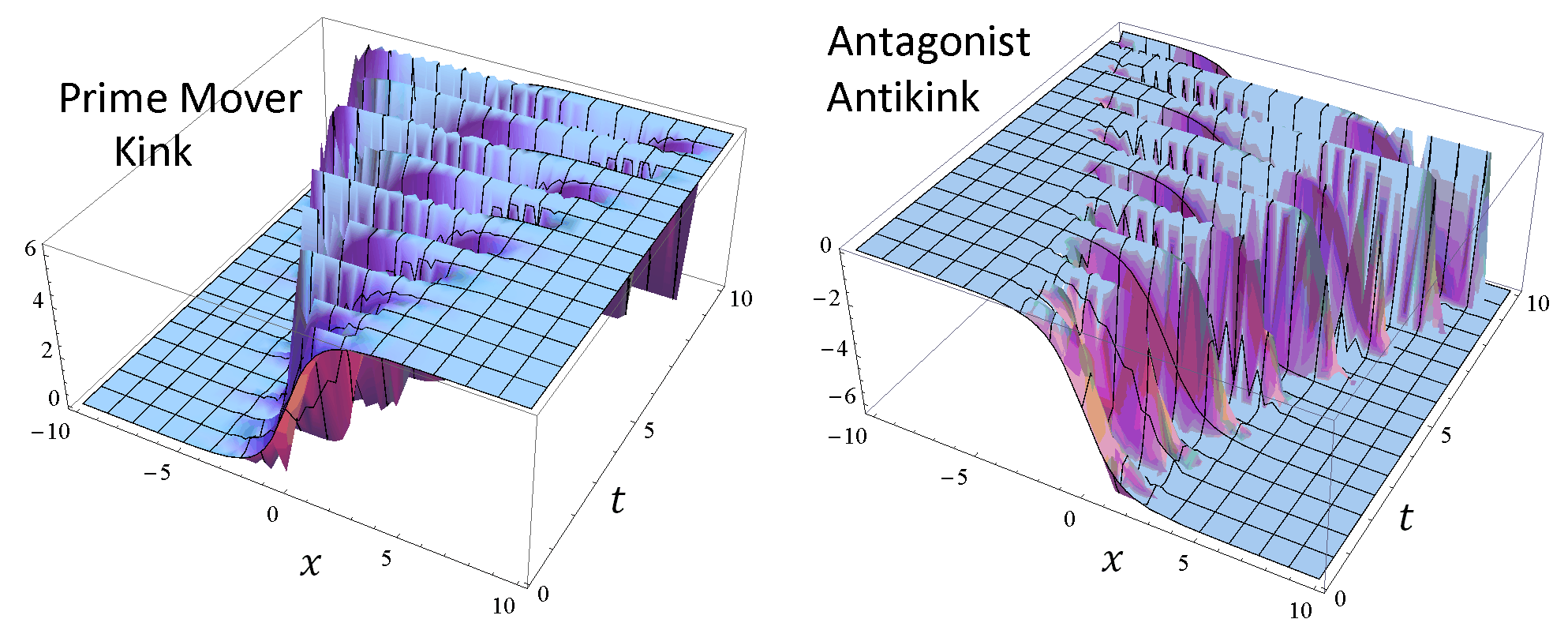}}
\caption{The SG kink-antikink solution (\protect\ref{kink}) of the
Sine-Gordon equation (\protect\ref{SG}), in which the signal velocity $v=\sin (5t)$ represents the excitation frequency. Kink represents the excitation of the prime mover (quadriceps), while antikink represents the
corresponding inhibition of the antagonist (hamstrings).
Time is
in seconds, muscle's length-change (shortening/stretching) is in
millimeters, so that for both muscular groups (quadriceps and hamstrings) the subsequent contraction amplitude will be 2cm. Both the kink and antikink MAPs (i.e., excitations and
inhibitions) have the so-called \emph{incomplete-tetanus} type of
quasi-periodic behaviors determined by the frequency of bicycle riding. }
\label{kink-antikink}
\end{figure}

A recent discovery includes the third important protein filament, \emph{titin,} an elastic, spring-like filament that forms the connection bridge between the Z-disc and the myosin filament, thus playing a major role in the passive fiber force generation and the \emph{eccentric} muscular contraction (see, e.g. \cite{WisdomDelp} and the references therein).

The understanding  of muscular contraction on the molecular level began with the discovery and further research of myosin filament and its shortening role \cite{t1}. Shortly after that,  with improved technology came the realization that the process of muscular contraction was possible with, not only one, but two filaments: myosin and actin. The sliding filaments theory (published side-by-side in Nature by Andrew Huxley and Hugh Huxley \cite{HuxleyA,HuxleyH}) most accurately describes this pulling process. That same year Natori discovered the third protein in the sarcomere \cite{t5}, which was later published under the name `connectin' by Maruyama \emph{et al.} \cite{t6} and is today known as `titin'. Later A.\@ Huxley mathematically explained the sliding filament theory and (unintentionally) promoted it as the `cross-bridged theory' \cite{t4} (he had an active correspondence with the Science Editor who encouraged him to publish his mathematical contribution as the molecular theory of muscular contraction) \cite{t16}.

Parallel to the story of the three proteins involved in muscular contraction, there are three types of contraction. The concentric contraction perfectly fitted in the cross bridge theory and gave the microscopic foundation to classical Hill's concentric force-velocity relationship \cite{Hill1}. The isometric contraction was in similar way explained by Gordon (A. Huxley's co-worker) \cite{t8} in 1966. The eccentric contraction remained inappropriately explained  for a long time, until its important role in muscular contraction was realized; it was clearly noticeable through the force-length relationship curve and its connection with the sarcomere, myosin and actin \cite{t16}. A prolific work by Herzog put the light into mysterious explanation of eccentric contraction \cite{t9}-\cite{t17} and the role of titin, through active and passive stretching of the sarcomere, and its bond with calcium during passive stretching.
The protein titin attracted the attention of scientific community, which resulted in the studies about its mechanical properties \cite{t18}, its `complex structure-dependant elasticity' \cite{t19}. Titin's palindromic assembly was formulated (titin-richest structures in the body are sarcomeres of striated and cardiac muscles \cite{t20}; titin is the largest protein molecule in the body, with the role in control of structure and extensibility of sarcomeres and the source of actin stiffness \cite{t21}).

\section{Kink-Soliton Model for Excitations and Contractions of the Knee
Extensors and Flexors}

In this section, we will first present a kink excitation/inhibition model
for the antagonistic pair of knee muscles, then we will present the simplest
possible soliton contraction/relaxation model for the same pair of knee
muscles, and finally we will couple them into a working kink-soliton model
of excitation-contraction, to be subsequently used to model bicycle riding
movement. For general overview of solitons and kinks, see \cite{Ablowitz}.
The coupled excitation-contraction model will be used to generate the
normalized force-time relation for various muscular length changes. In the
next section, this force-time relation will be used to generate the
resulting torque in the knee joint, that is, to control the
extension-flexion cycle in the human knee.

\begin{figure}[h]
\centerline{\includegraphics[width=16cm]{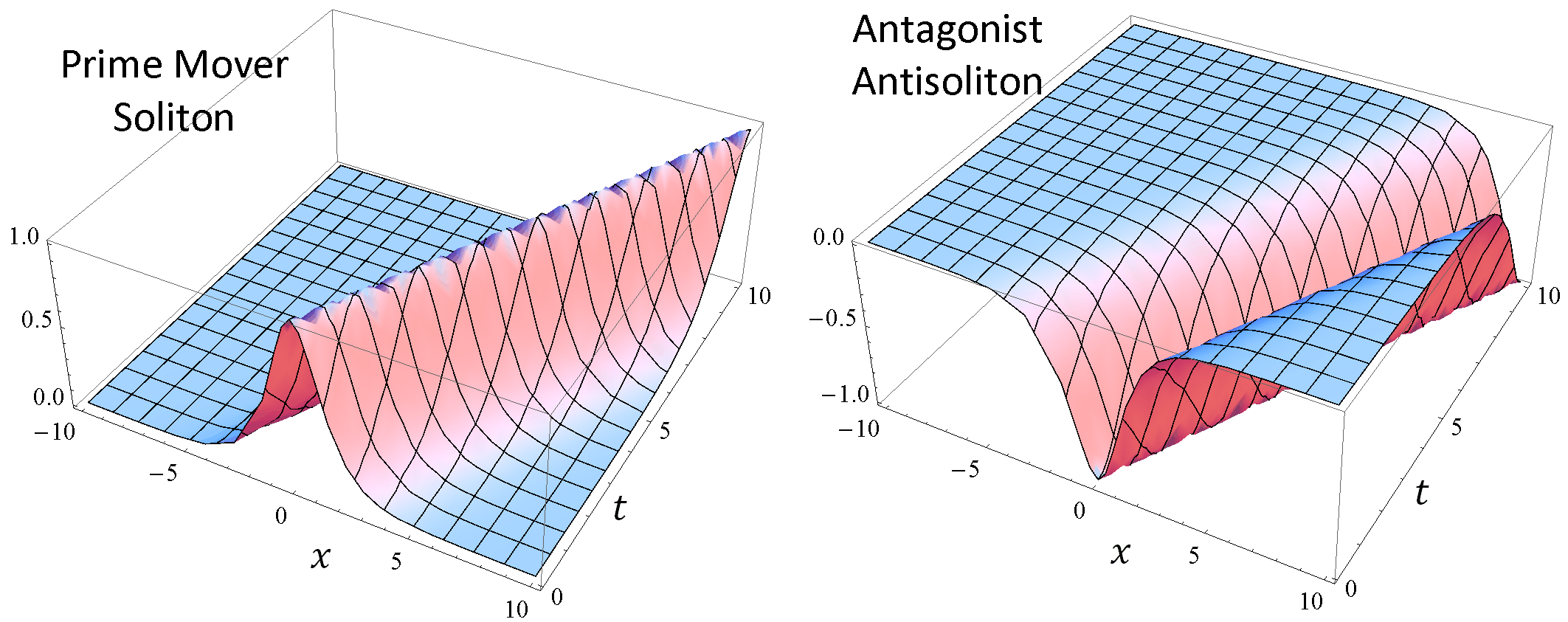}}
\caption{The soliton-antisoliton solution (\protect\ref{sech}) of the KdV
Eq. (\protect\ref{KdV}) with $\protect\tau=0.1s$ with the generated force
amplitude normalized to 1. Again, time is in seconds, muscle's length-change
is in millimeters, giving the total contraction amplitude of 2cm for both muscles. Soliton represents the normalized contraction of the
prime mover (quadriceps), while antisoliton represents the
corresponding normalized relaxation of the antagonist (hamstrings).}
\label{soliton-antisoliton}
\end{figure}

\subsection{Isolated Sine-Gordon Excitations/Inhibitions of the Knee
Actuators}

To start with the solitary modeling of muscular excitations and
contractions, recall that a thermal wave alternative to the classical,
Nobel-awarded, Hodgkin-Huxley electrical neural-conduction theory \cite%
{H-H,Hodgkin} was proposed a decade ago by \cite{HJ1,HJ2}, which was
recently generalized in \cite{ivSG,pulse}, where various forms of the
\textit{neural action potential} have been represented by a variety of
solitons, kinks and breathers of the (1+1) \textit{Sine-Gordon} (SG)
equation:
\begin{equation}
u_{xx}=u_{tt}+\sin u,  \label{SG}
\end{equation}%
where $u=u(x,t)$ is the real spatiotemporal function (a scalar field) and
indices denote partial derivatives (e.g., $u_{xx}\equiv \partial
^{2}u/\partial x^{2}$).

\begin{figure}[h]
\centerline{\includegraphics[width=16cm]{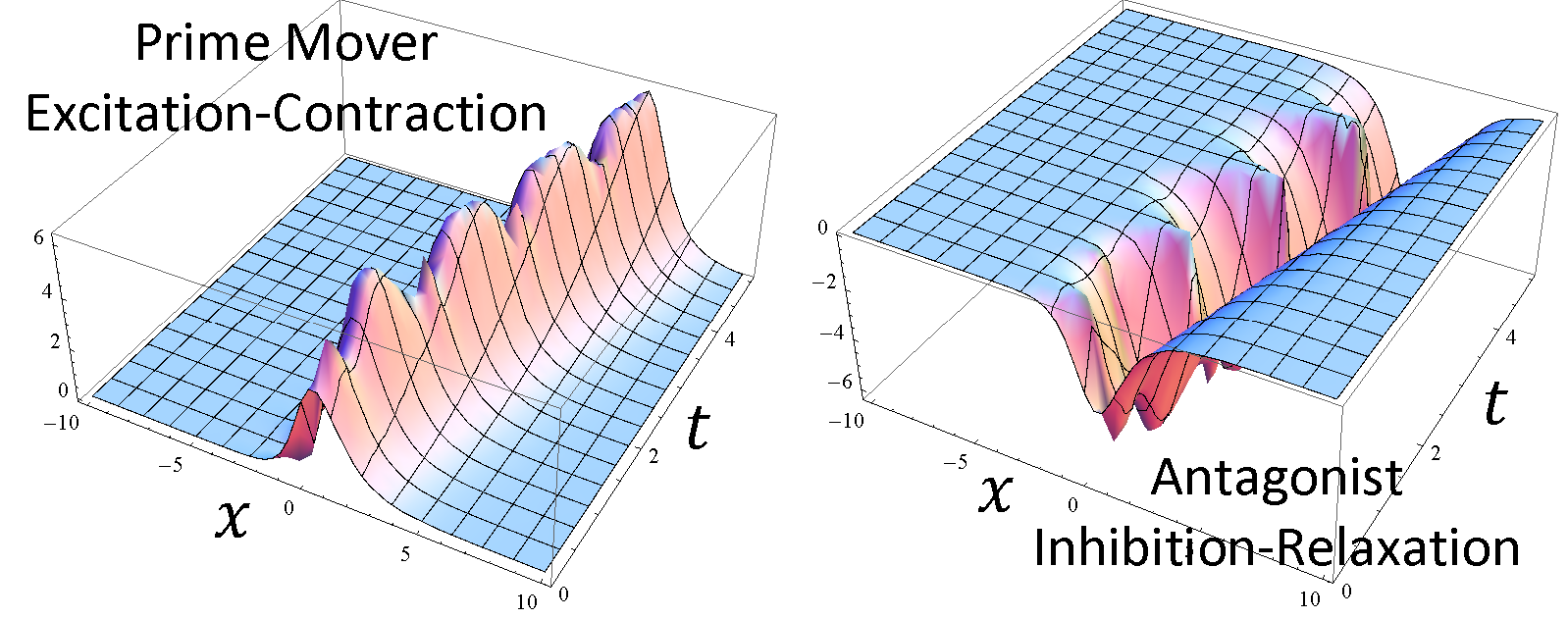}}
\caption{The kink-soliton model (\protect\ref{coupling}) for the
excitation-contraction coupling (left) with the corresponding
inhibition/relaxation coupling (right) of the knee extensors (quadriceps) and
flexors (hamstrings), representing the normalized muscular force-length-time
relation. Again, time is in seconds, muscle's length-change is in
millimeters, giving the contraction amplitude of 2cm for both muscles.}
\label{kink-soliton}
\end{figure}

Using the fact that muscles are excitable tissues similar to nerves (only
about 100 times slower), we can apply Eq. (\ref{SG}) to model various forms
of \textit{muscular action potential} (MAP), or muscular excitation, by
interpreting the driving term $\sin u$ as either induced internally by
neural stimuli, or externally by electro-muscular stimulation (see \cite%
{TijSpr14} and the references therein).

To describe the \emph{excitation/inhibition} action of a pair of mutually
antagonistic knee muscles (i.e., quadriceps femoris and hamstrings groups)
as the actuators in the knee joint, a particularly useful solution of Eq. (%
\ref{SG}) is the following SG kink-antikink pair (see Figure \ref%
{kink-antikink}):
\begin{equation}
u(x,t)=4\tan ^{-1}\left[ \pm \exp \left( \frac{x-tv}{\sqrt{1-v^{2}}}\right) %
\right] ,  \label{kink}
\end{equation}%
where the positive sign corresponds to the prime mover's kink and negative
sign to the antagonists's antikink, and $v$ represents the
excitation/inhibition signal velocity that is a function of the neural (or,
electro-muscular) stimulation frequency. For example, if we use the SG-kink
MAP to describe the simultaneous excitation of the left quadriceps and the
right hamstrings, then the SG-antikink MAP corresponds to the inhibition of
the left hamstrings and the right quadriceps.

\subsection{Isolated Korteweg-de Vries Contractions/Relaxations of the Knee
Actuators}

Next, we come to the isolated of muscular
contractions/inhibitions. To model the \emph{titin-influenced actomyosin contractions,} we recall that solitary models of muscular
contractions were pioneered by Davydov in \cite{DavydovSolit} and
subsequently elaborated in \cite{GaneshSprSml}. These models, representing
oscillations of peptide group Amid I inside spiral structures of myosin
filaments, can be represented by the solutions of the \textit{Korteweg-de
Vries} (KdV) equation:
\begin{equation}
u_{t}+uu_{x}+u_{xxx}=0.  \label{KdV}
\end{equation}%
Out of a variety of known \textsl{sech}, \textsl{tanh} and \textsl{cnoidal}
solutions of the KdV Eq. (\ref{KdV}), to model the \emph{%
contraction/inhibition} action of the knee extensor-flexor pair (with the
normalized amplitude), we start with the simplest solution (see \cite%
{Ablowitz}) in the form of the following soliton-antisoliton sech-pair:
\begin{equation}
u(x,t)=\pm \,\mathrm{sech}(x-t+\tau).  \label{sech}
\end{equation}%
where the positive sign corresponds to the prime mover's contraction soliton
and negative sign to the antagonist's inhibition antisoliton, and $\tau $
represents the time-delay between the previous excitation and the current
contraction (see Figure \ref{soliton-antisoliton}).

In particular, the positive sech soliton $u(x,t)=\mathrm{sech}(x-t+\tau)$ can be interpreted as a normalized 3D model for the muscular force-length relation (see Figures 6 and 7 in \cite{WisdomDelp}).

\subsection{Kink-Soliton: Excitation-Contraction Coupling of the Knee
Actuators}

Now we are able to define the whole \emph{excitation-contraction coupling}\footnote{The excitation-contraction coupling is an electro-chemical process through which the MAP triggers contraction of muscle fibers. Biochemically, it is performed by the release
of Ca$^++$ ions from the sarcoplasmic triads, their binding by troponin to remove the blocking action of tropomyosin, with the simultaneous rearrangement and exposition of the actin filaments and subsequent activation of myosin cross-bridges, to start the actin-myosin sliding fueled by the ATP hydrolysis (see, e.g. \cite{GaneshSprSml,GaneshWSc} and the references therein).} of the prime mover, together with the corresponding inhibition/relaxation coupling in the antagonist,
as a (normalized) product of the SG kink-antikink (\ref{kink}) and the KdV
soliton-antisoliton (\ref{sech}):
\begin{equation}
u(x,t)=\pm \tan ^{-1}\left[ \exp \left( \frac{x-tv}{\sqrt{1-v^{2}}}\right) %
\right] \,\mathrm{sech}(x-t+\tau ).  \label{coupling}
\end{equation}%
The kink-soliton model (\ref{coupling}) is depicted in Figure \ref%
{kink-soliton}. It is a spatiotemporal model of the normalized muscular
force-length-time relation describing both (left and right) quadriceps and
hamstrings muscles during the bicycle pedaling.

\section{Hill-Type Synergetic Control of the Knee Extension-Flexion Cycle}

\begin{figure}[h]
\centerline{\includegraphics[width=9cm]{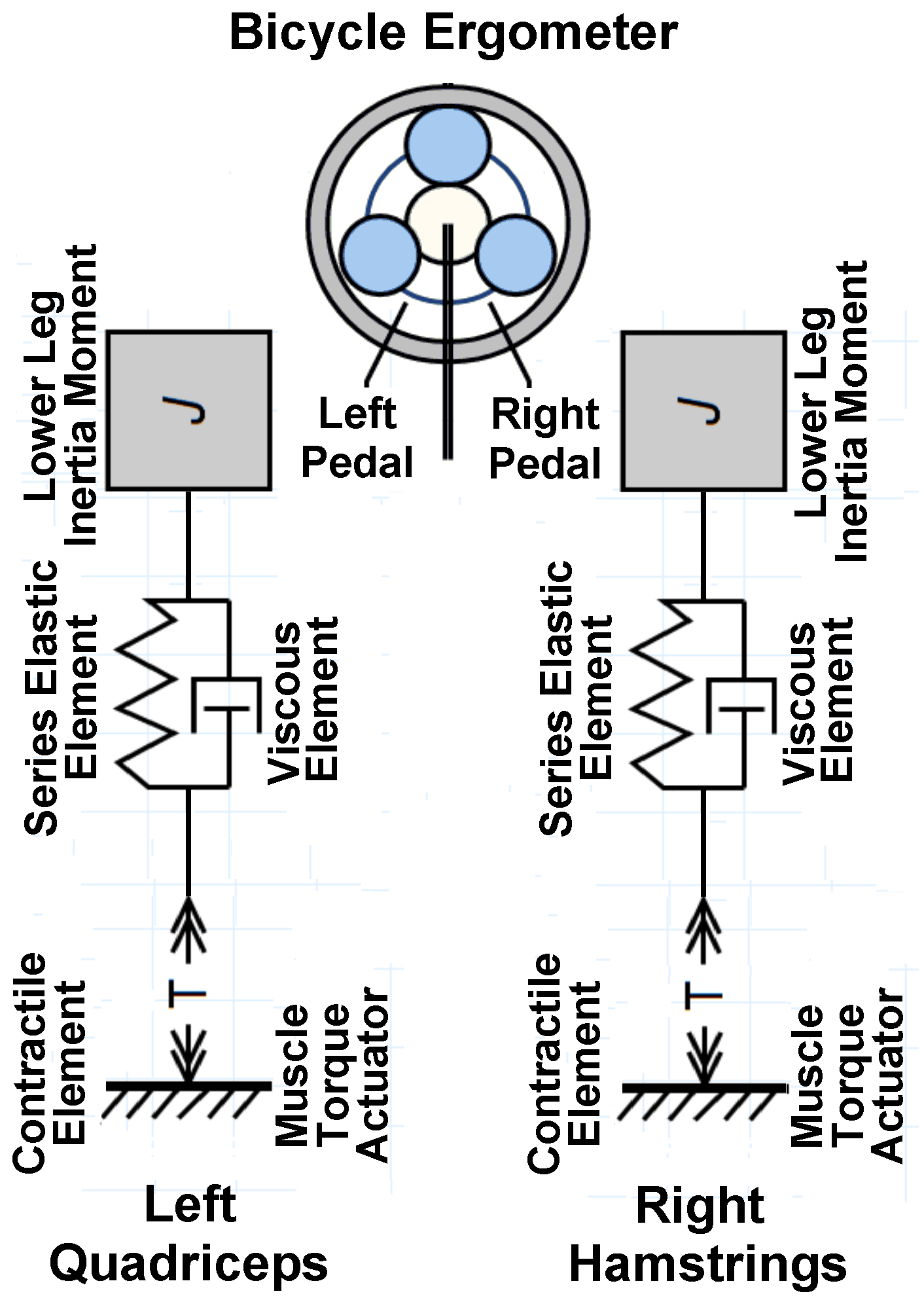}}
\caption{A synergy of two angular Hill-type muscular models driving the bicycle.}
\label{angHill}
\end{figure}

\subsection{Three-Component Joint Model Driven by the Kink-Soliton
Excitation-Contraction}

Now we can model the extension-flexion cycle for both knees during the bicycle driving movement (see Figure \ref{angHill}), including the
left quadriceps-driven extension, with the joint angle $q_{L}^{Q}(t)$ and
the angular velocity $\dot{q}_{L}^{Q}(t),$ as well as the right
hamstrings-driven flexion, with the joint angle $q_{R}^{H}(t)$ and the
angular velocity $\dot{q}_{R}^{H}(t)$. For this, the classical Hill's%
\footnote{%
Nobel Laureate A.V. Hill was the father of both muscular mechanics and
exercise physiology.} three-component muscle model (with contractile, series
elastic and damping elements \cite{Hill1,Hill2,Hill3}) is implemented using
nonlinear spring, linear damper and the kink-soliton excitation-contraction. In particular, the cubic spring is used, within the framework of Hill's series elastic component, to represent both the internal muscular elasticity of \emph{titin} (see Figure 2 in \cite{WisdomDelp}) and the external musculotendon complex of quadriceps and hamstrings.

\begin{figure}[h]
\centerline{\includegraphics[width=16cm]{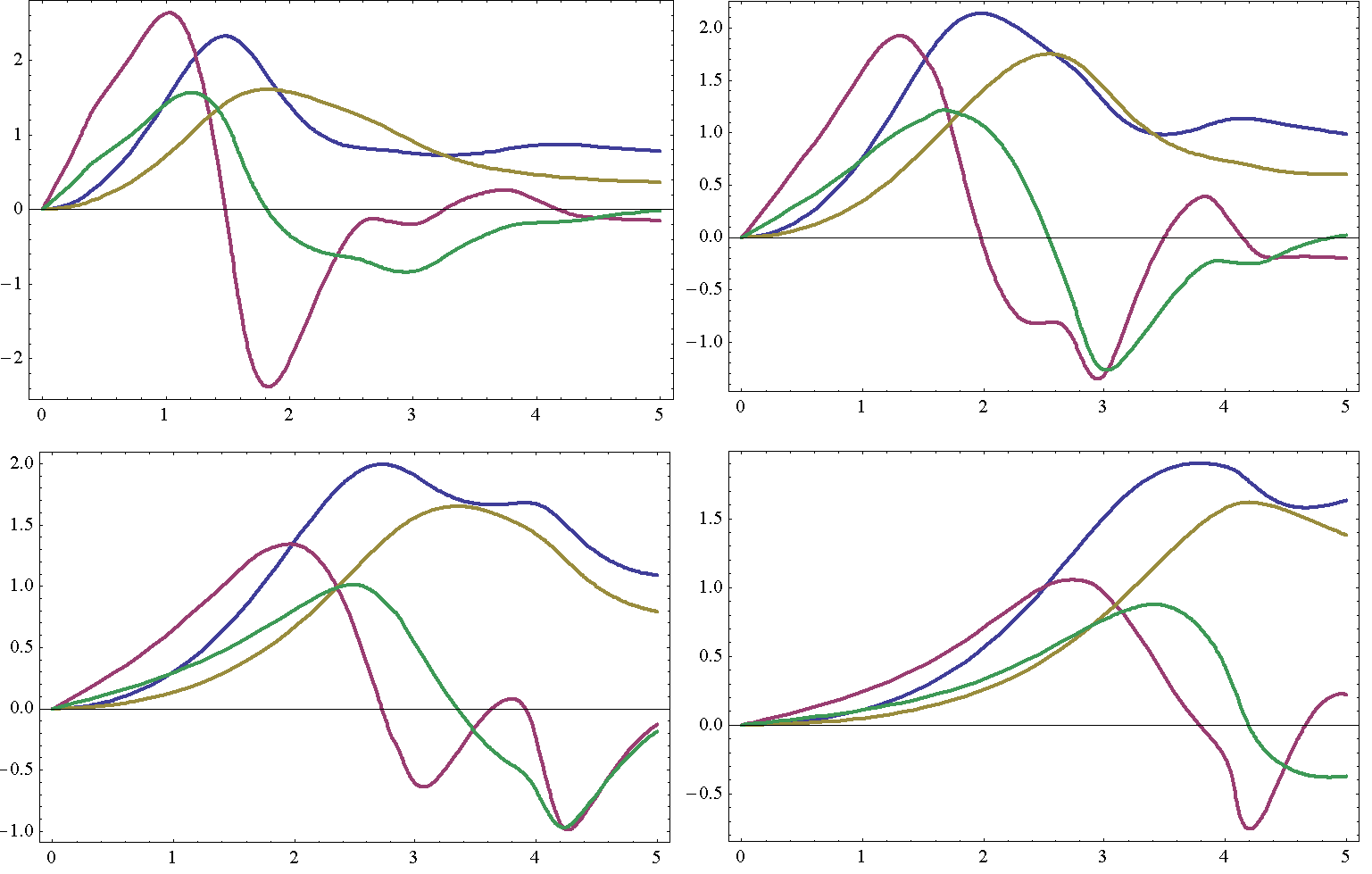}}
\caption{A 10s long simulation of the extension-flexion cycle for both knees defined by the equations of motion (\protect\ref{Hill}) and driven by muscular torques (\protect\ref{torque}), for both quadriceps and hamstrings length-changes (from stretched to contracted state) of $x=1,...,4$ centimeters. Angles $q(t)$ are in radians, while angular velocities $\dot{q}(t)$ are in $rad/s$.}
\label{Ode4}
\end{figure}

Assuming, for simplicity, the constant joint lever-arm for the knee joint
and consequently, the rotational joint form of Hill's model is formulated (see
Figure \ref{Ode4}) using the following two inertia-spring-damper equations
of rotational flexion-extension motion in the knee joint:
\begin{eqnarray}
\text{\textrm{Left knee extension}} &\mathrm{:}&J\,\ddot{q}_{L}^{Q}(t)+a\,%
\dot{q}_{L}^{Q}(t)+b\,q_{L}^{Q}(t)^{3}=c\,T(t,x),\quad \left( \mathrm{for\,\
}x\text{=}1,...,8;~\tau \text{=}0.1\right) ,  \notag \\
\text{\textrm{Right knee flexion}} &:&J\,\ddot{q}_{R}^{H}(t)+a\,\dot{q}%
_{R}^{H}(t)+b\,q_{R}^{H}(t)^{3}=\frac{c}{2}T(t-\tau ,x),  \label{Hill} \\
\text{\textrm{driven by the torque}} &\text{:}&T(t,x)= \tan ^{-1}\left[ \exp
^{\left( \frac{x-tv}{\sqrt{1-v^{2}}}\right) }\right] \,\mathrm{sech}%
(x-t+\tau ),  \label{torque}
\end{eqnarray}%
with normalized values of all included muscular parameters (lower leg's
moment of inertia $J$, Hill's angular spring $a$, damper $b$ and torque
amplitude $c$): $J$=$a$=$b$=$c$=$1$.\footnote{Hamstrings is naturally significantly weaker than quadriceps (which is an anti-gravitational muscle). This difference/disbalance is increased in athletes because of the squats-with-weights exercises for the quadriceps. This situation causes frequent hamstrings injuries among sprinters (and football/soccer players), partly because under fatigue hamstrings relaxation is slower than quadriceps contraction and partly because of the shoes with pins that put even more stress on the hamstrings.}

\subsection{Qualitative Model Validation: Force-Velocity Relation}

\begin{figure}[h]
\centerline{\includegraphics[width=16cm]{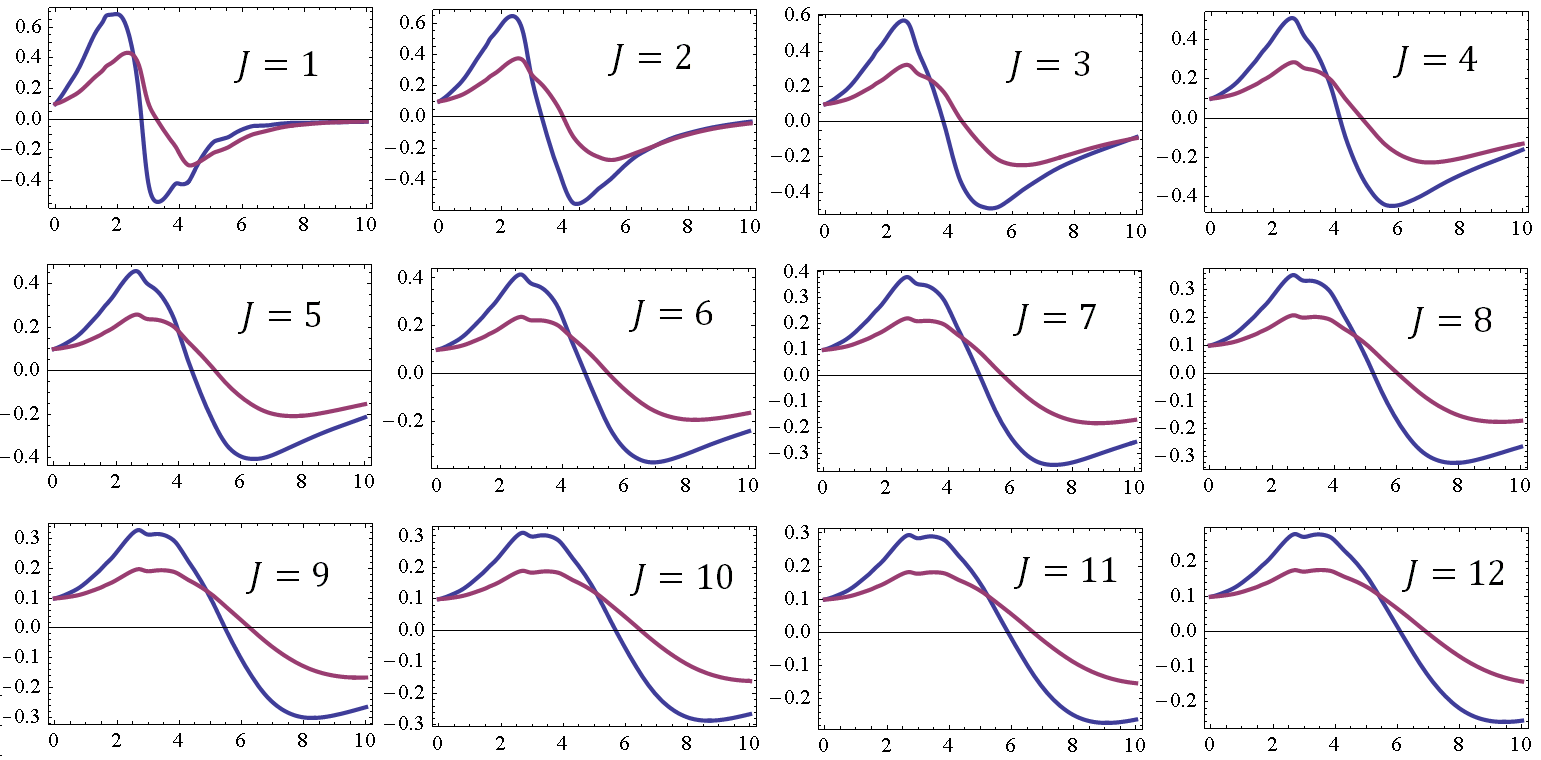}}
\caption{Angular velocities $\dot{q}(t)$ (in $rad/s$) simulated from Eq. (\protect\ref{Hill}) and Figure \ref{Ode4}, for both quadriceps and hamstrings length-change of $x=2$cm, with increasing (normalized) inertia moment $J\in[1,12]$. We can see that (angular) velocity in both muscles reduces with increased (inertial) loading, which qualitatively agrees with Hill's force-velocity relation.}
\label{f-v}
\end{figure}

\begin{figure}[h]
\centerline{\includegraphics[width=16cm]{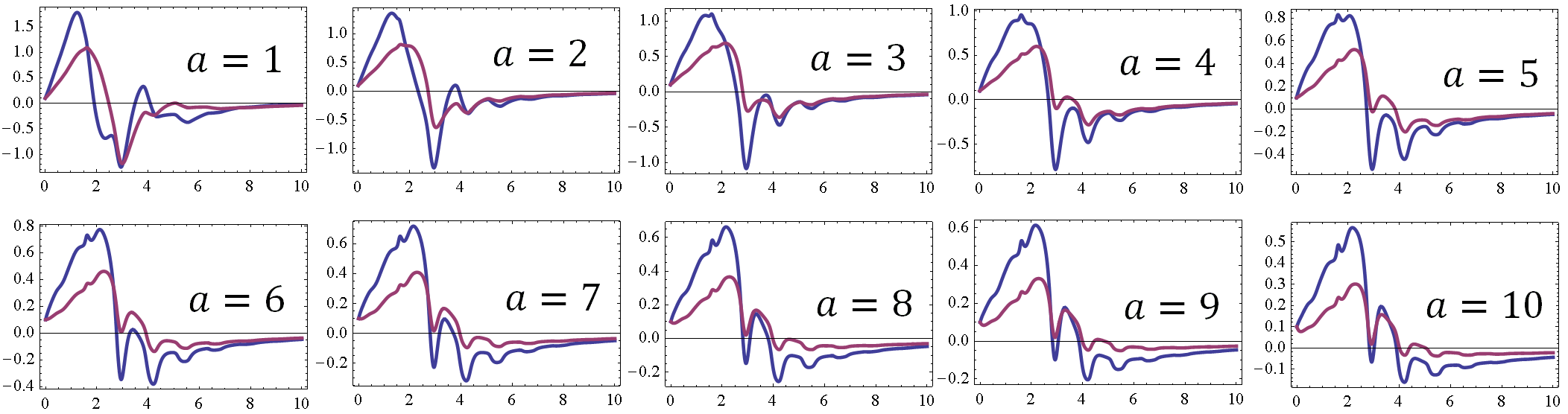}}
\caption{Angular velocities $\dot{q}(t)$ (in $rad/s$) simulated from Eq. (\protect\ref{Hill}) and Figure \ref{Ode4}, for both quadriceps and hamstrings length-change of $x=2$cm, with increasing (normalized) friction coefficient $a\in[1,10]$. We can see that (angular) velocity in both muscles reduces with increased (frictional) loading, which qualitatively agrees with Hill's force-velocity relation.}
\label{f-v2}
\end{figure}

Finally, the basic test for any biomechanical model is Hill's fundamental force-velocity relation. Figures \ref{f-v} and \ref{f-v2} show that our normalized model qualitatively satisfies force-velocity relation, in the sense that increased loading, either inertial or frictional, reduces the speed of movement. We could expect that with putting the correct biomechanical values of all included parameters, instead of the currently used values normalized to 1, would produce a proper hyperbolic shape of Hill's relation. However, for the purpose of the present theoretical model, the qualitative agreement suffices.

For more biomechanical details on bicycle riding (which are not at the focus of
the present work), see e.g. \cite{NeptuneCycling} and the references therein.

\section{Application to the Implantation of the Knee Joint
Endoprosthesis}

From surgical perspective of the \emph{total knee arthroplasty} (TKA), the successful implantation of the \emph{knee joint endoprosthesis,} i.e., placing the component in the optimal position, so to ensure both \emph{stability} and \emph{mobility} of the joint by femoral and tibial alignment, requires solid knowledge of anatomy of lower limbs and complete knee biomechanics, as well as a full respect to neural, vascular and soft-tissue structures (see \cite{Gojko1} and the references therein).

 Recovery after the TKA requires specialized outpatient physical therapy protocols, including intensive strengthening and functional exercises, like riding the bicycle ergometer, combined with carefully designed aquatic functional exercises performed in the non-gravitational posture. Based on patient progress, the intensity of both land-based and aquatic exercises should be gradually increased until optimal recovery is reached (see \cite{Pozzi} and the references therein).

\section{Conclusion}

In this theoretical paper we have presented three fundamental force relations: (i) force-length, (ii) force-time and (iii) force velocity - which span all levels of muscular mechanics, from the single sarcomere, to the whole body biomechanics. However, we remark here that there is the fourth important biomechanical relation, the \emph{force-angle relation,} which modulates the above three relations in real situations of human movement. For example, any athlete who is doing squat-with weights exercises knows that if they can lift the maximal weight of 100 kg from the deep squat posture, they can lift over 200 kg from the half-squat posture, thus showing that the quadriceps muscle is the strongest when the knee is almost extended. The same principle holds for hamstrings, as well as for triceps and biceps in the elbow joint -- all these muscles are the strongest when the leg or arm is almost extended.

It is important to acknowledge the latest research done about the human fascia \cite{Guimberteau} and how this research connects to both knee modelling and its subsequent tendons and ligaments \cite{Hagemeister}. Fascia is present in every single one of our 650 skeletal muscles. From microscopic levels to macroscopic levels, it is within every physiological process our bodies make. As an extension of the previous information; muscular contraction proteins and their interactions with surrounding tissue can also explain the connection fascia has with degenerative changes in joints (i.e., rheumatoid and osteoarthritis) as well as its connection to the immune and lymphatic systems. Discoveries of these papers are examples of an evolutionary approach to scientific research. These findings in vivo processes allow experiments, which are beneficial for humans, to be done without the need to involve any animal testing. With that said we believe this is the future of ethical research procedures.

\end{document}